\documentclass[reprint,prl,showpacs,superscriptaddress]{revtex4-1}

\usepackage{amssymb}
\usepackage{amsmath}    
\usepackage{graphicx}   
\usepackage{verbatim}        
\usepackage{color}      

\begin{document}

\title{Electron dynamics, $ \gamma $ and $ e^{-}e^{+} $ production by colliding
laser pulses}

\author{M. Jirka}
\affiliation{Institute of Physics of the CAS, ELI-Beamlines Project, Na Slovance
2, 182 21 Prague, Czech Republic}
\affiliation{Faculty of Nuclear Sciences and Physical Engineering, Czech
Technical University in Prague, Brehova 7, 115 19 Prague 1, Czech Republic}
\author{O. Klimo}
\affiliation{Institute of Physics of the CAS, ELI-Beamlines Project, Na Slovance
2, 182 21 Prague, Czech Republic}
\affiliation{Faculty of Nuclear Sciences and Physical Engineering, Czech
Technical University in Prague, Brehova 7, 115 19 Prague 1, Czech Republic}
\author{S. V. Bulanov}
\affiliation{Japan Atomic Energy Agency, Kansai Photon Science Institute, 8-1-7
Umemidai, Kizugawa, Kyoto, 619-0215 Japan}
\author{T. Zh. Esirkepov}
\affiliation{Japan Atomic Energy Agency, Kansai Photon Science Institute, 8-1-7
Umemidai, Kizugawa, Kyoto, 619-0215 Japan}
\author{E. Gelfer}
\affiliation{National Research Nuclear University MEPhI, Kashirskoe shosse 31,
Moscow, 115409 Russia}
\author{S. S. Bulanov}
\affiliation{Lawrence Berkeley National Laboratory, Berkeley, California 94720,
USA}
\author{S. Weber}
\affiliation{Institute of Physics of the CAS, ELI-Beamlines Project, Na Slovance
2, 182 21 Prague, Czech Republic}
\author{G. Korn}
\affiliation{Institute of Physics of the CAS, ELI-Beamlines Project, Na Slovance
2, 182 21 Prague, Czech Republic}

\begin{abstract}
The dynamics of an electron bunch irradiated by two focused colliding 
super-intense laser pulses and the resulting  $\gamma $ and $e^{-}e^{+} $ 
production are studied. 
Due to attractors of electron dynamics in a standing wave created by colliding 
pulses the photon emission and pair production, in general, are more efficient 
with linearly polarized pulses than with circularly polarized ones.
The dependence of the key parameters on the laser intensity and wavelength 
allows to identify the conditions for the cascade development and  $\gamma
e^{-}e^{+}$ plasma creation.
\end{abstract}


\maketitle

With the advent of $ 10~\mathrm{PW} $ laser facilities, the new and so far 
unexplored field of ultra-intense laser matter interaction will become 
accessible experimentally \cite{ELIwww}.
The intensities of the order of $ 10^{23-24}~\mathrm{W/cm^{2}} $ will be
achieved in these interactions, therefore the possibility of efficient 
generation of gamma-ray photons or even electron-positron pairs has attracted 
much attention in the last decade (see review article \cite{ADiP-2012} and 
references therein). 
In a strong electromagnetic field, electrons can be accelerated to such high 
energy that the radiation reaction starts to play an important role 
\cite{RAD}. 
Moreover, a new regime of the interaction can be entered, dominated by quantum 
electrodynamics (QED) effects such as pair production and cascade development 
\cite{ADiP-2012, Bell2008}. 
If a photon with sufficient energy is emitted due to multiphoton Compton 
scattering and then interacts with $ n $ laser photons, new electron-positron 
pair can be created via the multiphoton Breit-Wheeler process \cite{Breit1934}. 
Since the probabilities of the photon emission and pair creation depend on the 
particle momentum, on the electromagnetic field strength, and on their mutual 
orientation, it is necessary to elucidate the motion of electrons (positrons) 
in the electromagnetic field in the strong radiation reaction regime.

In this Letter we present the analysis of the electron motion and photon  
emission modeled as a discreet process in the electromagnetic (EM) standing 
wave (SW) generated by two colliding focused short super-intense laser pulses  
interacting with an electron bunch. 
The interaction of charged particles with an intense EM fields is characterized 
by two dimensionless relativistically invariant parameters 
\cite{Ritus1985}. 
First parameter is $a_0=eE_0/m_e\omega_0 c$, the dimensionless EM field 
amplitude, which measures the energy gain of an electron over the field 
wavelength in units of $2\pi m_ec^2$. 
It is often referred to as the classical nonlinearity parameter. 
Here $e$ and $m_e$ are the electron charge and mass, $E$ and $\omega_0$ are EM 
field strength and frequency, $c$ is the speed of light, respectively.
The second parameter is $\chi_e=[|(F_{\mu\nu}p_{\nu})^{2}|] ^{1/2}/m_{e}cE_{S} $
($\chi_\gamma=[|( F_{\mu\nu}\hbar k_{\nu})^{2}|] ^{1/2}/m_{e}cE_{S}$), where
$E_{S}=m_{e}^{2}c^{3}/e\hbar\simeq 1.3\times 10^{18}~\mathrm{V/m}$
\cite{Schwinger1951}, $ \hbar $ is the Planck constant, 
and $F_{\mu\nu}$ is the EM field tensor. 
The parameter $\chi_{e,\gamma}$ characterizes the interaction of electrons 
(positrons) and photons with the EM field. 
Depending on the energy of charged particles and field strength the interaction 
happens in one of the following regimes parametrized by $a_0$ and 
$\chi_{e,\gamma}$: (i) $a_0>1$, the electron dynamics is relativistic; 
(ii) $a_0>\epsilon_{{\rm rad}}^{-1/3}$, the interaction becomes radiation
dominated; (iii) $a_0>(2\alpha/3)^2\epsilon_{{\rm rad}}^{-1}$ (or $\chi_e>1$), 
the quantum effects begin to manifest themselves \cite{NIMA2011, diPiazza2010, 
Bulanov2015}; and (iv) $\chi_e>1$, $\chi_\gamma>1$, the interaction leads to an 
avalanche \cite{Fedotov2010, Bulanov2010, Elkina2011, Nerush2011}, the 
exponential growth of the electron-positrons and photons number. 
Here $\epsilon_{{\rm rad}}=4\pi r_e/3\lambda$ is the parameter indicating the 
strength of radiation reaction effects, $r_e=\alpha\hbar/m_e c$ is the 
classical electron radius.

We perform two-dimensional (2D) simulations of electron dynamics in a standing 
wave with the {\sf EPOCH} code \cite{Ridgers2014}, based on Particle-in-Cell 
(PIC) method. 
Photon emission and $e^-e^+$ pair creation via the Breit-Wheeler process 
\cite{Breit1934} are modeled using Monte-Carlo method. 
Two Gaussian laser pulses moving along the $x$-axis collide head-on in the 
center of the simulation box with the size of $40\lambda \times 40\lambda $.
Each laser pulse has the wavelength of $\lambda=1~\mathrm{\mu m}$, period of 
$T=2\pi/\omega$, focal spot full width at half maximum (FWHM) of 
$w_{0}=1.5\lambda$, FWHM duration of $9T=30~\mathrm{fs}$, intensity of $ 
1.11\times 10^{24}~\mathrm{W/cm^{2}} $ (so that $ a_{0}=900 $). 
We consider two cases of the laser polarization. The electric field oscillates 
along the $y$-axis in linearly polarized (LP) pulses and rotates about the 
$x$-axis in circularly polarized (CP) pulses.
At the focal spot, $ x=y=0 $, a $1\lambda$ diameter cloud of electrons  is 
represented by 76,000 quasi-particles. 
Its density is $ 0.05n_{c} $, where $n_{c} = m_{e}{\omega}^{2}/4\pi e^{2}$ is 
critical density.
In a moving reference frame the electron cloud can be considered as a bunch of 
electrons.
The  mesh size is $ \lambda/100=10~\mathrm{nm} $, the time step is $\approx 
0.01T $. 

Colliding laser pulses create a transient EM SW. 
When they completely overlap, the electric field strength is zero at nodes, $ 
x= (2n+1)\lambda/4 $, and  maximum at antinodes, $x= n\lambda/2 $, where $n$ is 
integer.
The magnetic field nodes and antinodes are shifted by $\lambda/4$ with respect 
to the electric field.
In an ideal infinite SW, the electric and magnetic fields remain zero in their
respective nodes, while in their respective antinodes they synchronously 
oscillate along the polarization axis in the LP case or synchronously rotate 
about the $x$-axis in the CP case.

As seen in Fig. \ref{fig-1}, the electron cloud is strongly distorted by 
colliding laser pulses.
A part of electrons is expelled from the high-field region, mainly in the 
transverse direction. The observed asymmetry between up and down expelled 
electrons is due to the carrier-envelope phase difference of the laser pulses.
Another part of electrons concentrates between the electric field antinodes.
On each laser half-period, approximately one tenth of the electrons trapped 
near the electric field nodes are driven away in the transverse direction, Fig. 
\ref{fig-1} (near $ x=\pm \lambda $, $ y=-2\lambda $).
In the case of CP, most electrons trapped in the SW are strongly localized
near the electric field nodes $ x=\pm \lambda/4 $, Fig. \ref{fig-1}(a).
These electrons drift away from the focal spot in the transverse direction
having relatively low value of $\chi_{e}$ parameter, $\chi_{e}<0.05$.
Higher $\chi_{e}$ parameter ($\approx 0.5$) is achieved by escaping electrons
crossing few spatial periods of the SW.

In contrast to CP, in the case of LP electrons diffuse inside  the SW period
so that a significant number of electrons appear in the vicinity of the 
electric field antinodes, Fig. \ref{fig-1}(b,c).
These electrons typically acquire significantly higher $\chi_e$ parameter
($\gtrsim 1$) than in CP case. This leads to a more prolific emission of 
photons in the LP case.
The electron concentration near antinodes has been emphasized in Ref.
\cite{Gonoskov2014}. 

\begin{figure}[h!]
\centering
\includegraphics[width=8.6cm]{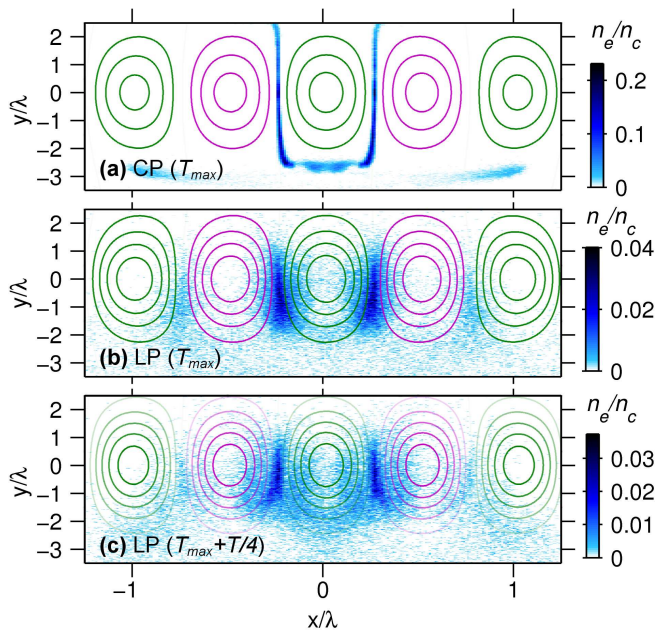}
\caption{\label{fig-1}
Electron density (colorscale, $ n_{e}/n_{c} $) in the $ (x,y) $ plane at $
T_{\max} $,
when the laser intensity is maximum in the focal spot, 
for different laser polarization: (a) CP, (b) LP, (c) LP at $ T_{\max}+T/4 $.
Green (magenta) curves for positive (negative) electric field $y$-component.}
\label{fig-1}
\end{figure}

\begin{figure}[h!]
\centering
\includegraphics[width=8.6cm]{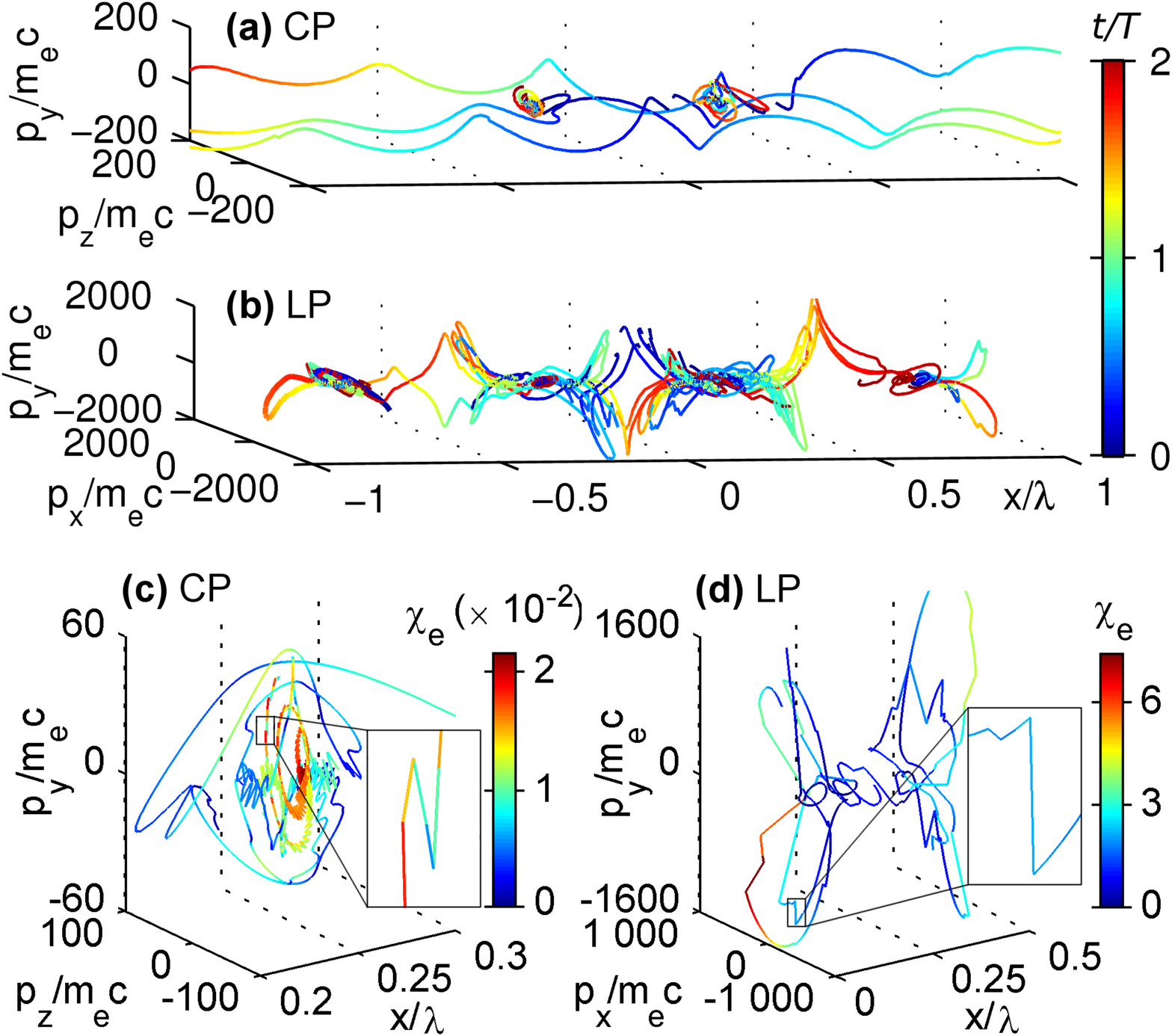}
\caption{\label{fig-2}
Electron trajectories in the phase subspace for $ T_{\max}<t<2T_{\max} $
for (a,c) CP and (b,d) LP. Colorscale represents (a,b) time $ t $ in laser
periods;
(c,d) the electron $ \chi_{e} $ parameter.}
\end{figure}

\begin{figure}[h!]
\centering
\includegraphics[width=8.6cm]{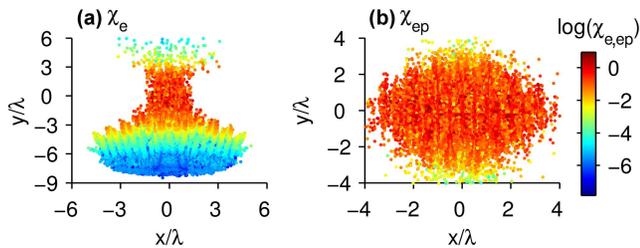}
\caption{\label{fig-3}
Parameters $ \chi_{e} $ and $ \chi_{ep}$ for seed  (a)
and Breit-Wheeler (b) electrons, respectively, in the $ (x, y) $ plane
at $ T_{\max} $  for LP. }
\end{figure}

\begin{figure}[h!]
\centering
\includegraphics[width=8.6cm]{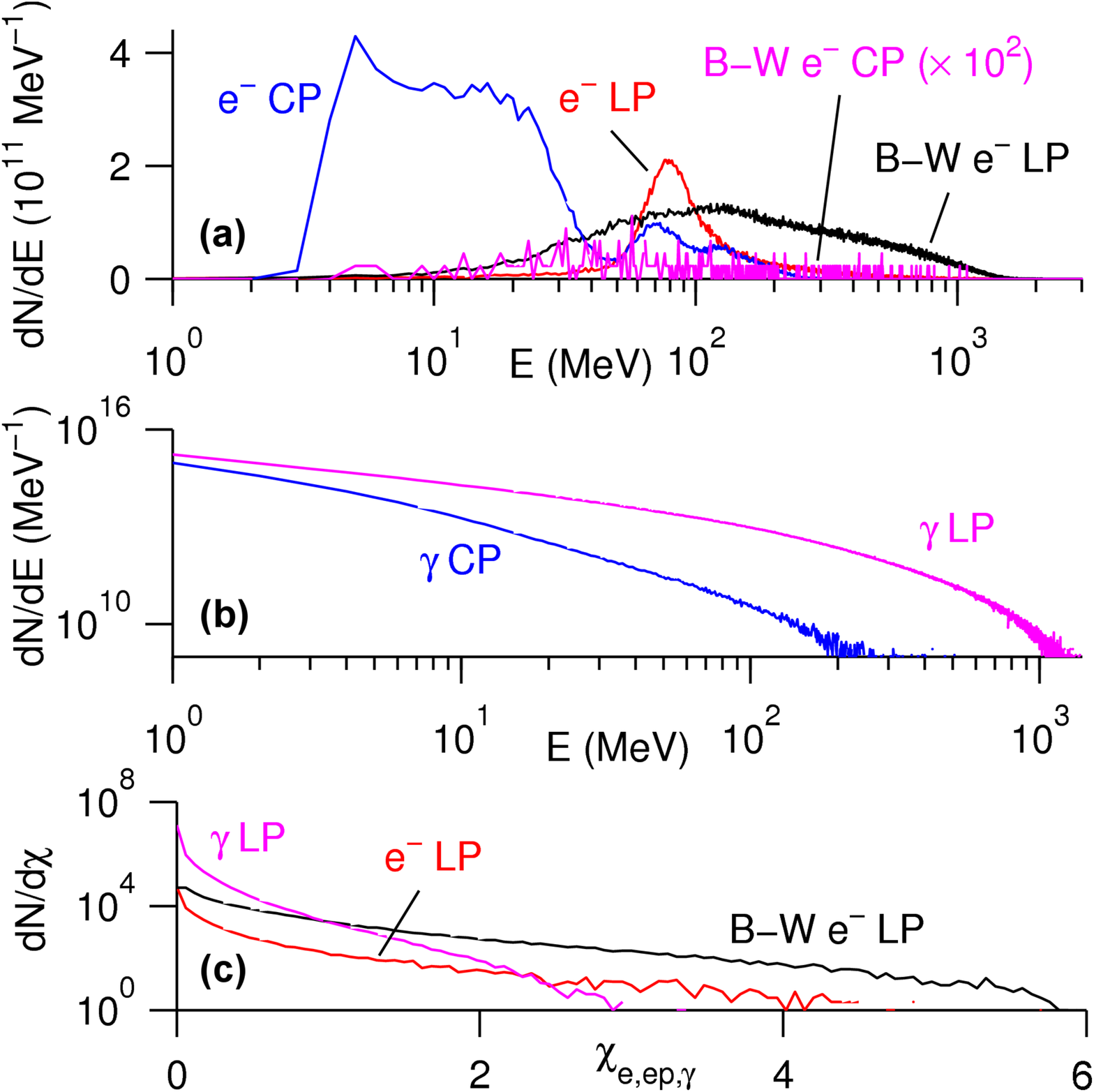}
\caption{\label{fig-4}
Energy spectra of seed and Breit-Wheeler electrons (a), and photons (b) at $
T_{\max} $ 
for different laser polarization.
(c) Electron and photon distribution with respect to $ \chi_{e}, \chi_{ep} $ and
$ \chi_{\gamma} $ 
parameters for LP.}
\end{figure}

Figure \ref{fig-2} shows representative electron trajectories in the phase
subspace $ (x, p_{y}, p_{z}) $ for CP (a,c) and $ (x, p_{x}, p_{y}) $ for LP 
(b,d), where $x$ and $p$ denote the electron longitudinal coordinate and 
momentum, respectively.
Trajectories colored with respect to time in frames (a,b) show that some 
electrons quiver near electric field nodes or oscillate in the half of the SW 
spatial period, while others escape transiting through many spatial periods of 
the SW along the $ x $-axis. 
In the CP SW, trapped electrons fall to the electric field nodes at 
$x=\pm\lambda/4$, Figs. \ref{fig-1}(a), \ref{fig-2}(a).
The electron rotation at the electric field antinode is strongly unstable, so 
that particles quickly depart from antinodes to nodes.
In the LP SW,  electrons appear also at the nodes at $ x=\pm3\lambda/4$, Figs. 
\ref{fig-1}(b,c), \ref{fig-2}(b). 

Fig. \ref{fig-2}(c,d) shows trajectories near the electric field node, colored 
according to the $\chi_e$ parameter.
The electron momentum and $\chi_e$ are much greater in the LP case than in the
CP case.
These trajectories show dynamic features of strange attractors at electric field
nodes and loops near antinodes predicted in Ref. \cite{Esirkepov2015} with a 
continuous emission model.
However, in this Letter the trajectories are not smooth, because of the quantum 
nature of the electron interaction with a high intensity EM field.
The photon emission makes a sudden change of particle momentum.
Since the probability of the multiphoton Compton effect corresponds to the 
emission of low energy photons in most cases, the electron momentum change is 
relatively small.
Being near an attractor, electrons after emission jump to another trajectory 
approaching the same or another attractor in the SW.
Therefore, the attractors characteristic to CP and LP SW (such as strange 
attractors and loops) found with a continuous emission model reveal itself in 
the case when the discrete quantum nature of emission dominates the dynamics.
In general, due to strongly dissipative dynamics, electrons ``forget'' their 
initial momentum and phase, therefore our conclusions concerning attractors 
remain valid also for high energy electron bunches, \cite{Esirkepov2015}.

An emitted photon  ballistically propagating through the SW creates an 
electron-positron pair via the Breit-Wheeler process. 
Since the photons emitted via the multistage Compton process have a power-law
spectrum, the number of Breit-Wheeler pairs is mainly determined by the number 
of photons whose radiation length is about the characteristic spatial extent of 
the EM SW \cite{Bulanov2013}. 
When these Breit-Wheeler pairs are born in the volume occupied by intense EM
field, they emit photons, thus increasing the photon number.
Moreover, $ \chi_{ep} $ parameter of newly created particles is higher than
$ \chi_{e} $ parameter of seed electrons for both CP and LP cases, Fig. 
\ref{fig-3}.
This is because Breit-Wheeler pairs are created near the EM field maxima.

Fig. \ref{fig-4}(a,b) shows the energy spectra of seed electrons, Breit-Wheeler 
electrons and photons for CP and LP.
Much more photons are produced in the LP case than in the CP case due to 
dynamic features seen in Figs. \ref{fig-2} and \ref{fig-3}.
Fig. \ref{fig-4}(c) gives the particle distributions with respect to $ 
\chi_{e}, \chi_{ep} $ and $ \chi_{\gamma} $ parameters for LP.
The number of Breit-Wheeler electrons is orders of magnitude greater than that
of seed electrons, while their ratio is almost constant for $ \chi_{e}, 
\chi_{ep} \gtrsim 1$. 
According to our 2D simulations, 10 Breit-Wheeler pairs are produced per 1 seed 
electron in the LP SW, indicating a strong avalanche. 
In the CP SW, this number is 650 times lower.

\begin{figure}[h!]
\centering
\includegraphics[width=8.6cm]{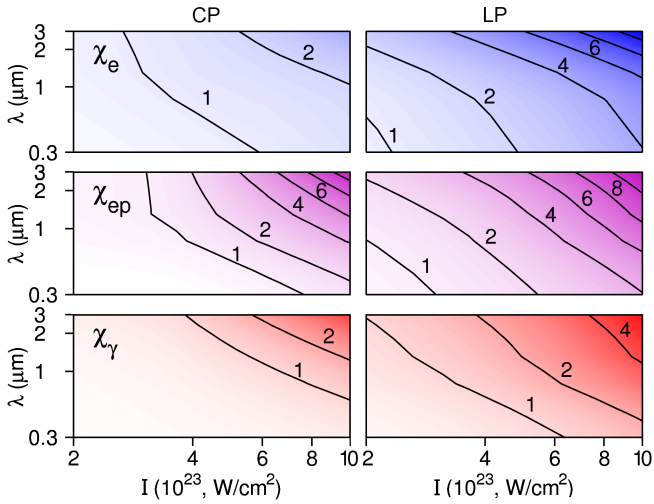}
\caption{\label{fig-5}
Maximum $ \chi_{e} $, $ \chi_{ep} $ and $ \chi_{\gamma} $ parameters
vs laser intensity, $I$, and wavelength, $\lambda$, for different laser
polarization.
}
\end{figure}

Different regimes of the electron bunch interaction with two super-intense 
colliding laser pulses, revealing a transition between the classical (radiation 
reaction force) and quantum (QED) description, are determined by the  laser 
wavelength and laser intensity \cite{Bulanov2015,Esirkepov2015}.
Considering the $ \chi_{\gamma} $ parameter of photons emitted in this 
interaction, one can determine whether the pair production is exponentially 
suppressed or a cascade develops.
For both CP and LP, we performed a set of simulations with the same parameters
as above, but for the laser wavelength in the range from $0.3~\mathrm{\mu m} $ 
to $3.0~\mathrm{\mu m} $ and the laser intensity varying from $ 
1.37\times10^{22}~\mathrm{W/cm^{2}}$ to $ 1.11\times10^{24}~\mathrm{W/cm^{2}}
$.
The maximum achieved $ \chi_{e}, \chi_{ep} $ and $ \chi_{\gamma} $ parameters
in the $(I,\lambda)$ space are presented in Fig. \ref{fig-5}.
Different regimes of interaction can be distinguished. 
While for $ \chi_{e}, \chi_{ep}  \ll 1 $ the radiation reaction is negligible, 
it is important for $ \chi_{e}, \chi_{ep} \gtrsim 1 $.
A cascade pair production starts when $ \chi_{\gamma} > 1 $.
This threshold is achieved with LP laser pulses for much lower intensity than
with CP ones. 
For a given intensity $ I $, higher values of $ \chi_{e} $ parameter are 
achieved as $ \lambda $ grows. 
As shown above, the $ \chi_{ep} $ parameter of Breit-Wheeler pairs is greater 
than the $ \chi_{e} $ parameter of seed electrons for both types of laser 
polarization. 
This difference is especially significant for CP laser pulses, as seen from Fig.
\ref{fig-5}.
Nevertheless, more intense CP laser pulses are needed to achieve $ \chi_{ep} = 
1$ in comparison with LP ones.
For a given laser intensity $ I $ and wavelength $ \lambda $, the values of $ 
\chi_{e}, \chi_{ep} $ and $ \chi_{\gamma} $ are always higher in the case of LP.
This is due to the different types of attractors of the dissipative electron 
dynamics in the CP and LP SW.
The presence of loops in the LP SW eases entering the QED dominated regime.

\begin{figure}[h!]
\centering
\includegraphics[width=8.6cm]{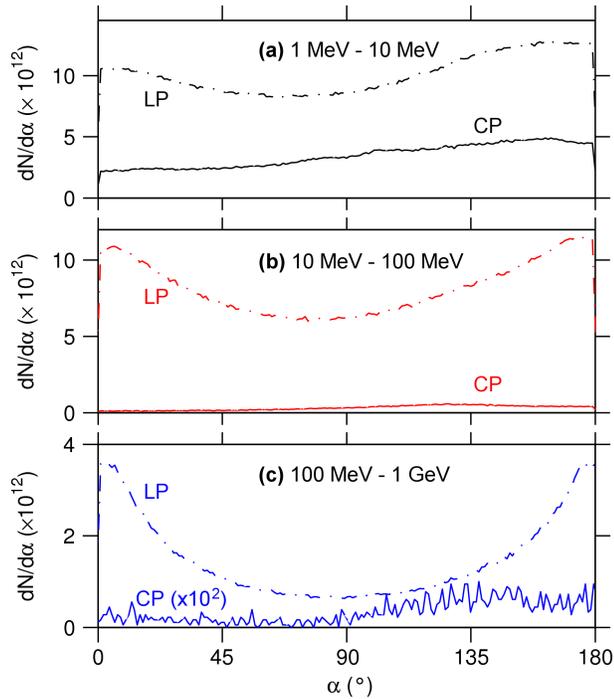}
\caption{\label{fig-6}
Angular energy distribution of photons having energy
(a) from 1 MeV to 10 MeV, (b) from 10 MeV to 100 MeV
and 
(c) from 100 MeV to 1 GeV. 
The angle $\alpha = 0^{o}$ ($\alpha = 90^{o}$) corresponds
to the $y$-axis ($x$-axis) direction.}
\end{figure}

In previous works concluding that the CP SW is more efficient for cascade
pair production than the LP SW, e.g., \cite{Nerush2011}, the electron (positron)
dynamics is considered exactly in the plane of the electric field antinode
on times shorter than the trajectory instability development. In contrast,
in our case particles have different initial positions, rearranged later by
colliding short laser pulses.
On the time-scale of few laser cycles particles fall into attractors 
characteristic to the dissipative system, even if particles initially 
move exactly in the electric field antinode plane. In addition, 
the recoil of photon emission at different angles quickly diffuses 
particles from the electric field antinodes.


The angular distribution of emitted photons also depends on the laser 
polarization as seen in Fig. \ref{fig-6}. 
In the case of LP, the photons are mainly emitted along the polarization 
direction being concentrated in the electric field antinode plane. 
In the case of CP, the photon emission is much lower, especially for higher 
photon energy.

In conclusion, in the interaction of an electron bunch with two colliding 
super-intense laser pulses, linearly polarized pulses ease entering the QED 
dominated interaction regime, facilitating a cascade development and $\gamma 
e^{-}e^{+}$ plasma creation. 
In contrast to circularly polarized laser pulses, linear polarization provides 
higher number of Breit-Wheeler pairs having higher energy, which gives much 
larger emitted photon number. 
The attractors of the electron motion in the electromagnetic standing wave
lead to more efficient photon emission in the linearly polarized standing wave.
Oscillating near loops electrons spend more time near electric field antinodes
thus producing more Breit-Wheeler pairs and emitting more high-energy photons. 
Our conclusions are valid for a configuration with obliquely colliding laser
pulses considered in a suitable reference frame.
We find that with an electron bunch interacting with two colliding linearly
polarized laser pulses having the intensity of $ 
5\times10^{23}~\mathrm{W/cm^{2}} $ and wavelength $1~\mathrm{\mu m} $, the
key parameters satisfy $ \chi_{e},\chi_{\gamma} \gtrsim 1 $.

Our work is supported by the project ELI: Extreme Light Infrastructure 
(CZ.1.05/1.1.00/02.0061) from European Regional Development Fund, 
Czech Science Foundation (M.J. 
and O.K., Project 15-02964S), Russian Foundation for Basic Research (E.G., 
Grant No. 13-02-00372), and U.S. DOE/SC (DE-AC02-05CH11231).
The {\sf EPOCH} code was developed under UK EPSRC grants EP/G054940/1, 
EP/G055165/1 and EP/G056803/1. We thank the MetaCentrum (LM2010005) and the 
CERIT-SC (part of the Operational Program Research and Development for 
Innovations,  CZ.1.05/3.2.00/08.0144).

\end{document}